\documentclass[11pt, a4paper, spanish]{article}
\usepackage[utf8]{inputenc}
\usepackage{empheq}
\usepackage{amsfonts}
\usepackage{graphicx}
\usepackage{hyperref}
\usepackage[normalem]{ulem}
\usepackage{epigraph}
\usepackage{bbm}
\usepackage{wrapfig}
\usepackage{lineno}
\usepackage{scalerel,stackengine}
\usepackage{physics}
\usepackage[hscale=0.75,vscale=0.8]{geometry}
\usepackage{amsmath, amsthm, amssymb, mathrsfs}
\usepackage{color}
\usepackage{authblk}
\usepackage[dvipsnames]{xcolor}
\usepackage{enumerate}
\usepackage{mathtools}
\usepackage{subcaption}
\usepackage{physics}

\usepackage[spanish]{babel}

\begin{document}

\title{Towards entropic uncertainty relations for non-regular Hilbert spaces}

\author{Alejandro Corichi}
\affil{Centro de Ciencias Matemáticas, Universidad Nacional Autónoma de México, UNAM Campus Morelia, A. Postal 61 3, Morelia, Michoacán 58090, Mexico}
\author{Angel Garcia-Chung}
\affil{ Departamento de Química, Universidad Autónoma Metropolitana, San Rafael Atlixco No. 186, Iztapalapa, Ciudad de México 09340, México}
\author{Federico Zadra}
\affil{  Department of Mathematics, University of Antwerp, Belgium   }

\date{\today}

\maketitle

\begin{abstract}
The Entropic Uncertainty Relations (EUR) result from inequalities that are intrinsic to the Hilbert space and its dual with no direct connection to the Canonical Commutation Relations. Bialynicky-Mielcisnky obtained them in \cite{bialynicki1975uncertainty} attending Hilbert spaces with a Lebesgue measure. The analysis of these EUR in the context of singular Hilbert spaces has not been addressed. Singular Hilbert spaces are widely used in scenarios where some discretization of the space (or spacetime) is considered, e.g., loop quantum gravity, loop quantum cosmology and polymer quantum mechanics. In this work, we present an overview of the essential literature background and the road map we plan to follow to obtain the EUR in polymer quantum mechanics.

\end{abstract}

\section{Introduction}

Information theory has emerged in the last decades as a consolidated branch of research permeating a vast number of disciplines, ranging from theoretical physics, chemistry, and statistics to artificial intelligence and data science \cite{shannon1948mathematical, cover1999elements, sen2011statistical, jaynes1957information, jaynes1957information2, fordbrandeis, esquivel2025information, deutsch1983uncertainty}. In physics and chemistry, a particular role is played by the Entropic Uncertainty Relations (EUR) \cite{bialynicki1975uncertainty} which constitute stronger conditions to characterize uncertainty in the quantum realm than the usual Heisenberg uncertainty relations (HUR). The EUR are based on the Shannon entropy -hence the term entropic- which, for a discrete probability distribution $\{p_j\}$, is defined as
\begin{align}
{\it H} = - \sum^N_j p_j \ln p_j, \label{discreteSE}
\end{align}
where the sum of $p_j \in [0,1]$ is normalized
\begin{align}
\sum^N_j p_j = 1,
\end{align}
and such that for an impossible event, $p_j = 0$, we impose
\begin{align}
\lim_{p_j \rightarrow 0} p_j \ln p_j : = 0.
\end{align}
The expression in (\ref{discreteSE}) characterizes the entropy of an information source \cite{shannon1948mathematical} and has been considered by some authors to have a more fundamental role in connection with statistical mechanics \cite{jaynes1957information, jaynes1957information2}. 
However, the Shannon entropy in (\ref{discreteSE}) is formulated in its discrete version. Its continuum expression\footnote{In position representation, indicated by the label $x$ and the index $p$ refers to the Shannon entropy in momentum representation.}
\begin{align}
H_x[\rho] = - \int \rho(\vec{x}) \, \ln \rho(\vec{x}) \, d\vec{x}, \label{continuumSE}
\end{align}
lacks, when applied to quantum mechanics, of an interpretation similar  to that in (\ref{discreteSE}). In (\ref{continuumSE}), $\rho(\vec{x})$ is a probability density which, in the case of a pure state takes the form
\begin{align}
\rho(\vec{x}) = | \Psi(\vec{x})|^2, \label{ProbDitri}
\end{align}
where $ \Psi(\vec{x}) \in {\cal H}$. Here, ${\cal H}$ is the Hilbert space used in what we call {\it the standard description of quantum mechanics} and it is given by
\begin{align}
{\cal H} = L^2(\mathbb{R}^n, d\vec{x}), 
\end{align}
where $n$ is the number of degrees of freedom of the system.

To clarify our point, notice that when working with discrete probability distributions, as in (\ref{discreteSE}), it is clear that the maximum entropy ${\it H}_{max} = \ln N$ acquires a direct meaning: the entropy takes its maximum value when all the probabilities are equal, implying that the system has no bias towards any preferred outcome. This interpretation is not possible in the case of a continuous probability distribution \cite{sen2011statistical} because the constant function with support in the entire real line is not an element of ${\cal H}$.

Another approach is to consider that Shannon entropy in (\ref{continuumSE}) measures how de-localized the wave function is, that is to say, it measures the spread of the wave function. However, there already exists a quantity doing this work: the standard deviation $\Delta x$ given by
\begin{align}
\Delta x = \sqrt{ \langle \Psi | \widehat{x}^2 | \Psi \rangle - \langle \Psi | \widehat{x} | \Psi \rangle^2}.
\end{align}
Hence, one might wonder what actually captures the Shannon entropy defined in (\ref{continuumSE}). 

The answer to these questions can be formulated using two arguments. First, the Shannon entropy in (\ref{continuumSE}) provides a measure of how spread the probability distribution (\ref{ProbDitri}) is in the whole real line, while the dispersion $\Delta x$ measures this spreading but around the mean value $\langle \Psi | \widehat{x} | \Psi \rangle$. In this regard, Shannon entropy is a more robust measure of the uncertainty of the wave function \cite{hirschman1957note}. 

The second argument goes hand in hand with the work of Iwo Bialynicki-Birula and Jerzy Mycielski in \cite{bialynicki1975uncertainty}. Bialynicki-Birula and Mycielski introduced the EUR, 
\begin{align}
H_x[\Psi] + H_p[\Psi]  \geq \ln(\pi \, e \, \hbar), \label{EURelation}
\end{align}
and proved that it constitutes a stronger version of the uncertainty relations compared to Heisenberg uncertainty relations (HUR), that is, one can derive the HUR using the EUR and not the contrary \cite{hirschman1957note, coles2017entropic}. As a result, a stronger version of the HUR, in the form of the EUR, has paved the way for an analysis of the uncertainty principle through new lenses \cite{coles2017entropic}.

It is also worth to point out that the EUR is derived without using the representation of the Canonical Commutation Relations (CCRs), but only referring to intrinsic features on the Hilbert space used and its dual under the Fourier transform. This feature captures our interest in the present project. Let us briefly present our main motivations in the next subsection.

\subsection{EUR for general representations of the CCs} \label{Motivation}

An important aspect of the EUR is that they were obtained for $L^p-$spaces with standard Lebesgue measures, $d\vec{x}$ and $d \vec{p}$. This yields a notable protagonism to a particular representation of the CCRs, the one given by
\begin{align}
\widehat{x}^a \, \Psi(\vec{x}) = x^a \, \Psi(\vec{x}), \label{PosRep} \\
\widehat{p}_b \, \Psi(\vec{x}) = \frac{\hbar}{\imath} \frac{\partial}{\partial x^b}\, \Psi(\vec{x}),  \label{MomRep}
\end{align}
and called {\it Schr\"odinger representation}. As a consequence, the expression for the EUR, when using any other representation, is an open question which has to be addressed and that has not been considered to the best of the authors' knowledge. 

On the other hand, one might argue that Stone-von Neumann's theorems \cite{rosenberg2004selective} guarantee that all regular representations of the CCRs are unitarily equivalent to the Schr\"odinger one in (\ref{PosRep} - \ref{MomRep}). Hence, there is no need to move towards the exploration of the EUR expression in other representations of the CCRs. But, as we will show, there are several reasons to study the expression for the EUR in non-Schr\"odinger representations.

Consider a Hilbert space given by ${\cal H}_G = L^2(\mathbb{R}, d \mu(x))$ where $d \mu(x)$ is a Gaussian measure of the form
\begin{align}
d \mu(x) = \frac{1}{\sqrt{\pi \, l^2}} \, e^{- (x/l)^2} \, dx,
\end{align}
which is normalized, 
\begin{align}
\int d \mu(x) = 1.
\end{align}
As a result of this normalization, the constant unit state $\varphi(x) = 1$ is an element of the Hilbert space ${\cal H}_G$. Notice that this state plays the role of a uniform-like distribution and this distribution, when considered in the discrete scheme, maximizes (\ref{discreteSE}). However, in this case, it is not clear which expression for the Shannon entropy is the appropriate one. For example, when using (\ref{continuumSE}), the Shannon entropy of $\varphi(x)$ is 
\begin{align}
H[\varphi] :=-  \int \varphi^2(x) \ln (\varphi^2(x)) \, d x = 0.
\end{align}
This contradicts the previous intuition, that is, the uniform {\it continuum distribution} has zero Shannon entropy. Moreover, the Fourier dual of $\varphi(x) = 1$ is a Dirac delta, $\widetilde{\varphi}(p) = \delta(p)$, and its Shannon entropy is clearly divergent. A direct consequence of these results is that no clear version of the EUR for these states using (\ref{continuumSE}) is available. Additionally, $\delta(x)$ is not in $L^p(d\mu)$, hence, the duality notion is also altered when a non-Lebesgue measure is considered.

Another argument is that representations in Hilbert spaces with Gaussian measures and configuration space $\mathbb{R}^n$, can be used to construct singular representations of the CCRs \cite{corichi2007polymer}. These singular representations are used in the context of quantizing the gravitational field \cite{rovelli2008loop, ashtekar2011loop, corichi2007hamiltonian} and also in scenarios where some discretization of the space affects the CCRs \cite{braunstein1996generalized}. Therefore, once the form of the EUR within a Hilbert space like ${\cal H}_G$ and its dual is obtained, we can expect that under certain limiting process we will obtain the uncertainty relations for singular representations of the CCRs.

Finally, when the configuration space is also different, say, given by the Schwartz space ${\cal S}$, and the measure is also Gaussian-like, then the EUR can be cast in the context of quantum field theory \cite{ashtekar1994algebraic, corichi2004schrodinger}. In this regard,  an important aspect when considering quantum field theory is whether the spacetime symmetries can be used to fix the quantum representation \cite{ashtekar1994algebraic}, such as on curved spacetimes. Therefore, extending the EUR to Hilbert spaces not only with Gaussian-like measures but also to different configuration spaces paves the way for the study of the relation between the EUR and the vacuum symmetries. 

Due to its relevance for our analysis, in Section \ref{EUR} we provide a summarized description of the derivation of the EUR given in \cite{bialynicki1975uncertainty}. In all these scenarios, the mathematical formalism developed to construct quantum representations, for finite and infinite degrees of freedom, shall be called {\it geometrical quantization}\footnote{This name is to differentiate it from the very well known geometric quantization formalism.}. For this reason, in Section \ref{GQ} we sketch the geometrical quantization procedure. In Section \ref{PQM} we introduce the main ingredients of the Polymer Quantum Mechanics scheme, which is used here as a representative of singular representations of the CCRs. Finally, in Section \ref{NS} we mention the next steps we plan to follow in order to carry out this research program.

\section{Entropic Uncertainty Relations} \label{EUR}

In this section, we describe the main steps followed in \cite{bialynicki1975uncertainty}. First, they considered a state $\Psi(\vec{x})$ in a $L^p(\mathbb{R}^n, d\vec{x})$ space. This space is formed by functions $\Psi: \mathbb{R}^n \rightarrow \mathbb{C}$ and such that its $p-$norm, $||\Psi ||_p$, defined as
\begin{align}
||\Psi ||_p := \left( \int \, |\Psi(\vec{x})|^p \, d\vec{x}  \right)^{1/p},
\end{align}
for a given $p\geq 1$.

The other element considered in \cite{bialynicki1975uncertainty} is the Fourier-dual state $\widetilde{\Psi}(\vec{p})$ in $L^q(\mathbb{R}^n, d\vec{p})$ which is given such that
\begin{align}
\widetilde{\Psi}(\vec{p}) = \frac{1}{(2 \, \pi \, \hbar)^n} \int  \, e^{- \frac{\imath}{\hbar} \vec{p} \vec{x}} \, \Psi(\vec{x}) \, d\vec{x}, 
\end{align}
and recall that this Fourier-transform is directly related with the fact that the momentum operator representation is of the form (\ref{MomRep}) since the plane waves are eigenfunctions of the momentum operator.

Here $q$ is the H\"older conjugate of $p$ given by
\begin{align}
q^{-1} + p^{-1} = 1,
\end{align}
and the next step is to relate the the $p$-norm of the state and the $q$-norm of its Fourier transform. Even though this step requires careful attention to technical mathematical details, it can be carried out by defining the $(p,q)$-norm~\cite{babenko1961inequality, beckner1975inequalities} as the smallest number $k(q,p)$ such that:
\begin{align}
\label{eqn:inequality}
||\widetilde{\Psi} ||_q \leq k(p, q) || \Psi ||_p 
\end{align}
for all state $\Psi$. Taking $q\geq 2$ and considering the H\"older condition, it turns out that
\begin{align}
k(p, q) = \left( \frac{2 \pi}{q} \right)^{n/2 q} \, \left( \frac{2 \pi}{p} \right)^{-n/2 p} \, \hbar^{\frac{n (2-q)}{2 q}} .
\end{align}

Once we have introduced all these elements, let us define the positive quantity
\begin{equation}
W(q) = k(p(q), q) || \Psi ||_{p(q)} - ||\widetilde{\Psi} ||_q ,
\end{equation}
which, according to Parseval-Plancherel theorem yields $W(2) = 0$. Moreover, Parseval-Plancherel theorem together with equation~(\ref{eqn:inequality}) implies that
\begin{align}
\left. \lim_{q \to 2^+} \frac{d W}{d q} \right\vert_{2^+} \geq 0,
\end{align}
with which, after inserting the expression for the right derivative we obtain
\begin{align}
H_x + H_p \geq n \, \ln (\pi \, e \, \hbar) \, N^2 - 4 N^2 \, \ln N \label{GEUR}
\end{align}
where $N = || \Psi ||_2$. If we take a normalized state $|| \Psi ||_2 = 1$, then (\ref{GEUR}) can be written as in (\ref{EURelation}), where $H_x$ and $H_p$ are the Shannon entropies for $\Psi(\vec{x})$ and $\widetilde{\Psi}(\vec{p})$ defined, respectively as
\begin{align}
H_x[\Psi] &:= - \int d\vec{x} \, |\Psi(\vec{x})|^2 \, \ln | \Psi(\vec{x})|^2, \\
H_p[\widetilde{\Psi}] &:= - \int d\vec{p} \, | \widetilde{\Psi}(\vec{p})|^2 \, \ln | \widetilde{\Psi}(\vec{p})|^2 .
\end{align}

According to \cite{bialynicki1975uncertainty}, the relation in (\ref{EURelation}) is a stronger version of the Heisenberg uncertainty relation due to it yields a stronger constraint. To check is, we can use Hirschman result \cite{hirschman1957note} (for $\hbar  = 1$) and state that
\begin{align}
H_x[\Psi] \leq \ln \sqrt{2 \pi e (\Delta x)^2} , \\
H_p[\widetilde{\Psi}] \leq \ln \sqrt{2 \pi e (\Delta p)^2},
\end{align}
from which we have that
\begin{align}
\frac{1}{2 \pi e} e^{H_x + H_p} \leq \Delta x \, \Delta p,
\end{align}
and after inserting (\ref{EURelation}) we obtain
\begin{align}
\frac{1}{2} \leq \Delta x \, \Delta p,
\end{align}
 which are the form of the familiar HUR.

 Let us now describe the main ingredients used in the formalism of geometrical quantization, which will be the core of our analysis when aiming at the form of the EUR for Gaussian or more general Hilbert spaces.

\section{Geometrical quantization} \label{GQ}

In this section we will sketch the main steps used in what we call the formalism of geometrical quantization. For simplicity we will consider a system with only one degree of freedom and we will focus on the kinematical description, i.e., no dynamics.

Consider the space $\Gamma = (\mathbb{R}^2, \Omega)$, where $\Omega$ is a symplectic structure
\begin{align}
\Omega = \left( \begin{array}{cc} 0 & 1 \\ -1 & 0 \end{array}\right),
\end{align}
and let us consider $\Gamma$ as a real linear space. We then introduce (and fix) a linear complex structure $J$, which is a map $J: \Gamma \rightarrow \Gamma$ such that $J^2 = - I$. Here $I$ is the identity matrix. It can checked that the more general form of $J$ is given by
\begin{align}
J = \left( \begin{array}{cc} a & b \\ - \frac{(1 + a^2)}{b} & -a\end{array}\right). \label{ComStr}
\end{align}
The eigenvalues of $J$ are $\pm \imath$, hence its eigenvectors $ \vec{u}_{\pm} \notin \Gamma$. The eigenvector with positive ($+\imath$) eigenvalue is called {\it positive frequency} eigenvector, while the other eigenvector is called the {\it negative frequency}.

We now complexify the space $\Gamma  \rightarrow \Gamma_{\mathbb{C}} $ and verify that every vector $\vec{X} \in \Gamma_{\mathbb{C}} $ can be written as
\begin{align}
\vec{X} = \vec{X}^+ + \vec{X}^{-},
\end{align}
where 
\begin{align}
\vec{X}^\pm = \frac{1}{2} \vec{X} \mp \frac{\imath}{2} J \vec{X}.
\end{align}

The space ${\cal S}^+$ given by
\begin{align}
{\cal S}^+ = \mbox{Span}\{ \vec{X}^+ \},
\end{align}
called {\it space of positive frequencies} will be used to construct what is called the {\it one particle Hilbert space}. To do so we first associate the inner product 
\begin{align}
\langle \vec{X}^+ | \vec{Y}^+ \rangle_J = - \imath ( \vec{X}^{+})^\dagger \Omega \vec{Y}^+
\end{align}
and construct the space $({\cal S}^+, \langle , \rangle_J )$ formed by positive frequency vectors. Then, Cauchy complete this space with the norm induced by this inner product. The Cauchy completion gives the Hilbert space ${\cal H}_J$. This Hilbert space is now used to construct the Fock spaces \cite{wald1994quantum}
\begin{align}
{\cal F}_B &= \mathbb{C} \oplus {\cal H}_J \oplus \left( {\cal H}_J \otimes {\cal H}_J \right)_s \oplus \left( {\cal H}_J \otimes {\cal H}_J \otimes {\cal H}_J \right)_s \oplus \dots \nonumber \\
{\cal F}_F &= \mathbb{C} \oplus {\cal H}_J \oplus \left( {\cal H}_J \otimes {\cal H}_J \right)_a \oplus \left( {\cal H}_J \otimes {\cal H}_J \otimes {\cal H}_J \right)_a \oplus \dots \nonumber
\end{align} 
where the index $a$ and $s$ stand for the antisymmetrization or the symmetrization of the tensor product. In the case of fermionic systems, the antisymmetric tensor product is the appropriate Fock space while for the case of bosonic systems is the symmetric one.

The quantum representation using this Fock-Hilbert space has been studied together with its unitary relation with the Schr\"odinger Hilbert space ${\cal H}_{Sch}$ \cite{corichi2004schrodinger, ashtekar1975quantum, corichi2002schrodinger}, in both flat and curved spacetimes for real scalar fields. At this point, the relevant observation is that in order to have a unitary relation between Fock representation and Schr\"odinger representation, we have to consider that the Schr\"odinger representation is given in a Hilbert space of the form
\begin{align}
{\cal H}_{Sch} = L^2({\cal S}, d\mu_{J})
\end{align}
where ${\cal S}$ is the Schwartz space in the case the real scalar field and $\mathbb{R}^n$ in the case of mechanical systems with $n$ degrees of freedom.  Concerning the measure, $d\mu_{J}$ depends on the parameter $b$ of the complex structure given in \ref{ComStr}. In the limit when $b$ goes to $0$ or to $\infty$, the Hilbert space used in polymer quantum mechanics is obtained \cite{corichi2007polymer}. As for the real scalar field, it is fixed using the symmetry of the spacetime, see \cite{corichi2004schrodinger, corichi2002schrodinger} for more details.

Let us now move to briefly describe the main ingredients of the polymer quantization which serves as a simplified example to explore singular representations.

\section{Polymer quantum mechanics} \label{PQM}

Consider a Hilbert given by ${\cal H}_{poly} = L^2(\mathbb{R}_d, dx_c)$ where $\mathbb{R}_d$ is the real line equipped with the discrete topology rather than the usual standard topology and $dx_c$ is the countable measure on it. Functions on this Hilbert space are given by linear combinations of discrete Kronecker deltas
\begin{align}
\Psi(x) = \sum_{ \{ x_j \} } \Psi_{x_j} \, \delta_{x, x_j}, \label{PolyPosState}
\end{align}
and such that the norm
\begin{align}
|| \Psi || = \sqrt{ \sum_{\{ x_j \}} |\Psi_{x_j} |^2 }< \infty
\end{align}
is finite.

This Hilbert space is a non-separable Hilbert space as can be seen from having an uncountable number of basis elements $\delta_{x, x_j}$, each one labeled by $x_j$, which is an arbitrary point over the real line. A direct consequence of this is that there are no possible infinitesimal translations and as a result, no momentum operator $\widehat{p}$ is possible.

To handle this situation, instead of working with infinitesimal translations, finite translation operators are used and are called the {\it holonomy} operator $\widehat{U}_\mu$ because they borrowed their names from the loop quantum gravity scheme. The parameter $\mu$ is a dimensionfull parameter with length units and, in the context of a discrete space\footnote{Another discretization of space different to that given by the topology of the configuration space $\mathbb{R}_d$} it is also assumed that there is a minimum length scale, denoted as $\mu^*$, whose value is  undetermined but fixed. This minimum scale implies that for any to space points $x_1$ and $x_2$ they satisfy the following relation
\begin{align}
(x_1 - x_2)/\mu^* \in \mathbb{Z},
\end{align}
i.e., they can be reached by a finite number of jumps of size $\mu^*$. 

With all these elements, it can be checked that the CCRs are now given by
\begin{align}
\left[ \widehat{x} , \widehat{U}_{\mu^*} \right] = - \mu^* \, \widehat{U}_{\mu^*},
\end{align}
which constitutes an example of modified CCRs \cite{braunstein1996generalized} and the representation of this modified CCRs is given as
\begin{align}
\widehat{x} \Psi(x) &= x \Psi(x), \\
\widehat{U}_{\mu^*} \Psi(x) &= \Psi(x + \mu^*).
\end{align}
An immediate consequence of this representation is that the Kronecker deltas are the eigenstates of the position operator, 
\begin{align}
\widehat{x} \, \delta_{x, x_j} = x_j \, \delta_{x,x_j},
\end{align}
thus, according to Born's postulate, the quantum particle is well localized. This is a crucial difference between this singular representation and the Schr\"odinger representation in (\ref{PosRep}, \ref{MomRep}) where the eigenstates of the position operator in (\ref{PosRep}) are Dirac deltas and as we know, they are not vectors in ${\cal H}$. On the other hand, no momentum operator exists in this singular representation but still, a Fourier transformation can indeed be defined as
\begin{align}
{\cal F}: {\cal H}_{poly} \rightarrow \widetilde{\cal H}_{poly}; {\cal F}[\delta_{x,x_j}] = e^{\frac{\imath}{\hbar} p x_j },
\end{align}
where $\widetilde{\cal H}_{poly}$ is the Fourier-dual space of ${\cal H}_{poly}$ and it is given by
\begin{align}
\widetilde{\cal H}_{poly} = L^2(\overline{\mathbb{R}}, dp_{Haar}).
\end{align}
Here, the configuration space $\overline{\mathbb{R}}$ is the Bohr-compactification of the real line\cite{velhinho2007quantum}. The states in $\widetilde{\cal H}_{poly}$ can be written as 
\begin{align}
\widetilde{\Psi}(p) = \sum_{ \{ x_j \} } \Psi_{x_j} \, e^{ \frac{\imath}{\hbar} p x_j}, 
\end{align}
where the coefficients are those given in (\ref{PolyPosState}) and the inner product when using the Haar measure $dp_{Haar}$ yields
\begin{align}
\langle e^{ \frac{\imath}{\hbar} p x_j} | e^{ \frac{\imath}{\hbar} p x_k} \rangle =  \delta_{x_j, x_k}.
\end{align}
Notably, the Fourier-dual of the position eigenstates are again the plane waves similar to what we have in the standard quantum mechanics but again, the plane waves are indeed elements of $\widetilde{\cal H}_{poly}$ whereas in the standard quantum mechanics they are not. 

Finally, it is worth to emphasize that Stone-von Neumann's theorems \cite{rosenberg2004selective} conditions do not hold for these singular representations. In particular the representation of the holonomy operator
\begin{align}
\langle \delta_{x,x_j} | \widehat{U}_{\mu} \delta_{x,x_j} \rangle = \langle \delta_{x,x_j} | \delta_{x,x_j - \mu} \rangle = \delta_{x_j, x_j - \mu},
\end{align}
 is not weakly continuous\footnote{For this analysis we have to consider an unfixed $\mu$.} which is one of the assumptions of the theorems. As a result, there is no unitary transformation that relates the physics obtained using this representation with the physics of the standard Schr\"odinger representation. To achieve some sort of equivalence, certain coarse graining procedures have to be implemented, see for example \cite{corichi2007hamiltonian}.

These are the main ingredients regarding the polymer quantum representation. Our main questions and the outline of the next steps will be described in the next section.

\section{Conclusions} \label{NS}

We have seen how different quantization procedures, summarized in Sections~\ref{GQ} and~\ref{PQM} lead to different Hilbert spaces, and how this difference relies on the measure used. In the setting of Geometrical Quantization, the induced Hilbert space is endowed with the Gaussian measure, while in the context of the polymer quantum mechanics, we are considering Haar and discrete measures.

Geometrical quantization allows us to connect the Fock-Hilbert space representation with the Gaussian-like Schr\"odinger representations. For the case of real scalar fields, this was done in 
\cite{corichi2004schrodinger, corichi2002schrodinger} but it has to be done for systems with finite degrees of freedom. Furthermore, in \cite{corichi2007polymer}, a connection between the Gaussian-like representation as described in subsection (\ref{Motivation}) was already given. This result was obtained by considering the two limits in which the parameters in the Gaussian measure go to zero (coordinate representation) or to infinity (momenta representation).

One of the most evident consequences in the ``changing'' measure is to obtain a different momentum operator. For example, we can indeed see that for a probability measure $d\mu(x)$ over $\mathbb{R}$, i.e. $\int_\mathbb{R} d\mu(x) = 1 $ such that:
\begin{equation}
    d\mu(x) = M(x) dx,
\end{equation}
with $M(x)\neq 0$ the more general momentum operator is given by:
\begin{equation}
 \widehat{p} =  \frac{ \hbar}{\imath} \frac{\partial}{\partial x} + i \frac{\partial \log(M(x))}{\partial x} + g(x),
\end{equation}
where $g(x)$ is a real function. 
We obtain these different forms since we require that the momentum 
operator is a first-order symmetric operator that satisfies the CCR.
Since in the geometrical quantization procedure the induced measure on the Hilbert space is related to the choice of the parameters defining the complex structure, we can also relate these parameters to the momentum operator.

The eigenvalue problem associated with the momentum operator induces a different version of the Fourier transform, which will be related to the EUR for Gaussian-measure Hilbert spaces. The modified Fourier transforms will be a gateway of the project and will lead to a challenge of the problem of extending the results in~\cite{babenko1961inequality, beckner1975inequalities}, and later to consequently extend the EUR reported in Section~\ref{EUR}.

We hope that this project helps shed light on the understanding of the EUR in more general contexts, and of course, we aim to present our findings in the next symposium: ``Applications of Information Theory in Natural Sciences''.

\section{Acknowledgements}
A.G.Ch would like to thank the CONAHCyT for a postdoctoral fellowship. F.Z. is partially founded by FWO-EoS project ``Beyond symplectic geometry''.


\end{document}